\documentclass[twocolumn, superscriptaddress,preprintnumbers, amsmath,epsfig, floatfix, prl]{revtex4}

\usepackage{graphicx}
\usepackage{dcolumn}
\usepackage{bm}
\usepackage{ulem}

\usepackage{times}

\usepackage[utf8]{inputenc}
\usepackage[english]{babel}
\usepackage{keyval}
\graphicspath{{figs/}{figs:}{\figs}}
\setkeys{Gin}{width=0.90\columnwidth}

\usepackage[usenames,dvipsnames]{color}
\usepackage[]{ulem}

\newcommand{\comment}[1]{\textcolor{red}{#1}}
\renewcommand{\comment}[1]{\relax}

\newcommand{\todelete}[1]{\textcolor{green}{\sout{#1}}}
\renewcommand{\todelete}[1]{\relax}

\begin{document}
\title{Surface-phase superconductivity in Mg-deficient V-doped MgTi$_2$O$_4$ spinel}
\date{\today}
\author{A. Rahaman}
\affiliation{Department of Physics, Indian Institute of Technology Kharagpur, Kharagpur - 721302, India}
\affiliation{Department of Physics, Yogoda Satsanga Palpara Mahavidyalaya, Palpara - 721458, India}
\author{T. Paramanik}
\affiliation{Department of Physics, Indian Institute of Technology Kharagpur, Kharagpur - 721302, India}
\affiliation{Department of Physics, School of Sciences, National Institute of Technology Andhra Pradesh, Tadepalligudem - 534102, India}
\author{B. Pal}
\affiliation{Department of Condensed Matter and Materials Physics, S. N. Bose National Centre for Basic Sciences, Block JD, Sector III, Salt Lake, Kolkata - 700106, India}
\thanks{Present Address of Dr. B. Pal: School of Physical Sciences, Indian Association for the Cultivation of Science, Jadavpur, Kolkata - 700032, India}
\author{R. Pal}
\affiliation{Department of Condensed Matter and Materials Physics, S. N. Bose National Centre for Basic Sciences, Block JD, Sector III, Salt Lake, Kolkata - 700106, India}
\author{P. Maji}
\affiliation{Department of Physics, Indian Institute of Technology Kharagpur, Kharagpur - 721302, India}
\author{K. Bera}
\affiliation{School of Nanoscience and Technology, Indian Institute of Technology Kharagpur, Kharagpur - 721302, India}
\author{S. Mallik}
\affiliation{School of Nanoscience and Technology, Indian Institute of Technology Kharagpur, Kharagpur - 721302, India}
\author{D. K. Goswami}
\affiliation{Department of Physics, Indian Institute of Technology Kharagpur, Kharagpur - 721302, India}
\author{A. N. Pal}
\affiliation{Department of Condensed Matter and Materials Physics, S. N. Bose National Centre for Basic Sciences, Block JD, Sector III, Salt Lake, Kolkata - 700106, India}
\author{D. Choudhury}
\email{debraj@phy.iitkgp.ac.in}
\affiliation{Department of Physics, Indian Institute of Technology Kharagpur, Kharagpur - 721302, India}

\begin{abstract}

\label{Superconductivity} Around fifty years ago, LiTi$_2$O$_4$ was reported to be first spinel oxide to exhibit a superconducting transition with highest T$_c$ $\approx$ 13.7 K. Recently, MgTi$_2$O$_4$ has been found to be the only other spinel oxide to reveal a superconducting transition with a T$_c$ $\approx$ 3 K, however, its superconducting state is realized only in thin film superlattices involving SrTiO$_3$. We find that a V-doped Mg$_{1-x}$Ti$_2$O$_4$ phase, which gets stabilized as a thin surface layer on top of stoichiometric and insulating V-doped MgTi$_2$O$_4$ bulk sample, exhibits high-temperature superconductivity with T$_c$ $\approx$ 16 K. The superconducting transition is also confirmed through a concomitant sharp diamagnetic transition immediately below T$_c$. The spinel phase of the superconducting surface layer is elucidated through grazing-incidence X-ray diffraction and Micro-Raman spectroscopy. A small shift of the sharp superconducting transition temperature ($\sim$ 4 K) with application of a high magnetic field (upto 9 Tesla) suggests a very high critical field for the system, $\sim$ 25 Tesla. Thus, V-doped Mg$_{1-x}$Ti$_2$O$_4$ exhibits the highest T$_c$ among spinel superconductors and also possesses a very high critical field.

\end{abstract}

\maketitle
Identification of new superconducting materials is an extremely fascinating and challenging task in the field of condensed matter physics. In this regards, spinel compounds, which are well known for exhibiting a plethora of functional properties due to a strong coupling between its charge, spin, orbital and lattice degrees of freedom, are rarely found to be superconductors. Around five decades ago (1967), some of the sulpho and seleno spinels were successfully synthesized with superconducting transition temperatures $\sim$ 4 K \cite{Nagata1995,Cava2013,Luo2022}. In the family of spinel oxides, superconductivity was first realized in
the mixed valent titanate spinel LiTi$_2$O$_4$ \cite{Johnston1973}, with the highest transition temperature (T$_c$) of 13.7 K \cite{Johnston1976}. While the mechanism driving the superconducting transition in LiTi$_2$O$_4$ still remains to be settled, the role of orbital degrees of freedom and spin-orbital fluctuations seem important \cite{Wu2004,Takeuchi2015,Jin2019,Moshopoulou1999,Jin2018}. Several investigations were performed to increase the superconducting T$_c$ of LiTi$_2$O$_4$ by doping at the Ti site with Mg, Mn, Li, Al, Cr ions, however, the T$_c$ was found to decrease rapidly with increase in doping percentage \cite{Edwards1988,Harrison1990,Goodenough1985,Yamauchi1994}. Superconductivity in the family of mixed titanate spinel oxide Mg$_2$TiO$_4$-MgTi$_2$O$_4$ remain controversial; in one group of studies, the Mg$_2$TiO$_4$-MgTi$_2$O$_4$ compounds were found to exhibit a zero resistive transition, albeit with the onset of diamagnetic signal at much lower temperatures (almost at 40 K smaller temperatures than the onset of zero resistance state) \cite{Irvine1991,Irvine1993,JTSIrvine1993,JIrvine1993}, other group of studies suggested these compounds to be instead semiconducting \cite{Yamauchi1994,Goodenough2005}. Recently, superconductivity has been reported in a superlattice consisting of MgTi$_2$O$_4$ and SrTiO$_3$ with a T$_c$ $\sim$ 3 K (where substrate induced strain was found to play a critical role) \cite{Jin2020} and in Mg: Ti$_9$O$_{10}$ (possessing orthorhombic Ti$_9$O$_{10}$ structure) film on the (011)-oriented substrate (MgAl$_2$O$_4$) with a T$_c$ $\sim$ 5 K \cite{Jin2022}. Bulk MgTi$_2$O$_4$, containing Ti$^{3+}$ ions, remains insulating (reported to be a Mott insulator \cite{LCraco2008,DiMatteo2004}) down to the lowest temperature and undergoes an insulator to high-temperature metal (or semiconducting \cite{Goodenough2005}) transition around 260 K. This phase transition is also accompanied with a Ti$^{3+}$-ion related Jahn-Teller distortion driven tetragonal to cubic structural and a Ti spin-singlet transition \cite{Schmidt2004,YUeda2002}. The low-temperature tetragonal phase hosts a unique tetramer orbital ordering involving the Ti t$_{2g}$ orbitals along $<$111$>$ direction and is chiral ($\it P$4$_1$2$_1$2) \cite{TMizokawa2005,LCraco2008,DiMatteo2004}. V doped MgTi$_2$O$_4$ still remains a Mott insulator down to the lowest temperature \cite{Sugimoto1998,Rahaman2021}, however, V doping leads to a unique mixed valence state for both Ti (Ti$^{3+}$ and Ti$^{4+}$) and V ions (V$^{3+}$ and V$^{2+}$), as it is energetically favourable for some of the Ti$^{3+}$ (3$\it{d}^1$) ions to donate their single electron (and thereby become Jahn-Teller inactive 3$\it{d}^0$) to the doped V$^{3+}$ (3$\it{d}^2$) ions (which also then acquire Jahn-Teller inactive 3$\it{d}^3$ configuration) \cite{Rahaman2021}. This mixed valence state of the transition metal ions in V-doped MgTi$_2$O$_4$ accompanied with a unique band structure leads to exotic functional properties, like a dc current-induced insulator to metal switching at ultra-low electric field \cite{Rahaman2021}. The present results on the emergence of superconductivity on the surface layer of V-doped MgTi$_2$O$_4$, further charge-doped due to Mg deficiency, with a much higher T$_c$ ($\sim$ 16 K) (the highest T$_c$ among spinels) and a very high upper critical magnetic field value is, thus, extremely promising.
\\To prepare the polycrystalline MgTi$_{1.4}$V$_{0.6}$O$_4$ sample, MgO (10\% excess Mg taken following \cite{Goodenough2005}), V$_2$O$_3$, TiO$_2$ and metallic Ti powders were thoroughly mixed, ground and casted into a pellet. The resultant pellet was subsequently annealed at 1080$^0$C under vacuum condition in a sealed quartz tube. While the bulk of the sample was found to be black in colour (corresponding to the MgTi$_{1.4}$V$_{0.6}$O$_4$ phase), a combination of two phases could be detected as a thin-surface layer, one of them being the black-coloured bulk phase and another an emergent grayish coloured phase. To investigate the structural phase of the surface layer, we have carried out the grazing-incidence X-ray diffraction (GIXRD) with very low incident angle using Cu-K$\alpha$ source. The powder X-ray diffraction (XRD) of the bulk sample was obtained after scraping off the thin grayish surface layer to investigate the structural phase. Micro-Raman experiments were carried out using a 532 nm laser source to further investigate the structural phases of the grayish and dark regions of the thin surface layer. Temperature-dependent four-probe resistivity and magnetization measurements were carried out using a Physical Property Measurement System (PPMS). The resistivity measurements were carried out by painting electrical contacts on the MgTi$_{1.4}$V$_{0.6}$O$_4$ sample, with and without (obtained by scrapping with a sand paper) the thin grayish surface layer, as shown in the insets of Fig.\ref{super}(a).
\begin{figure}[h!]
	\begin{center}
		\scalebox{1.15}
		{\includegraphics{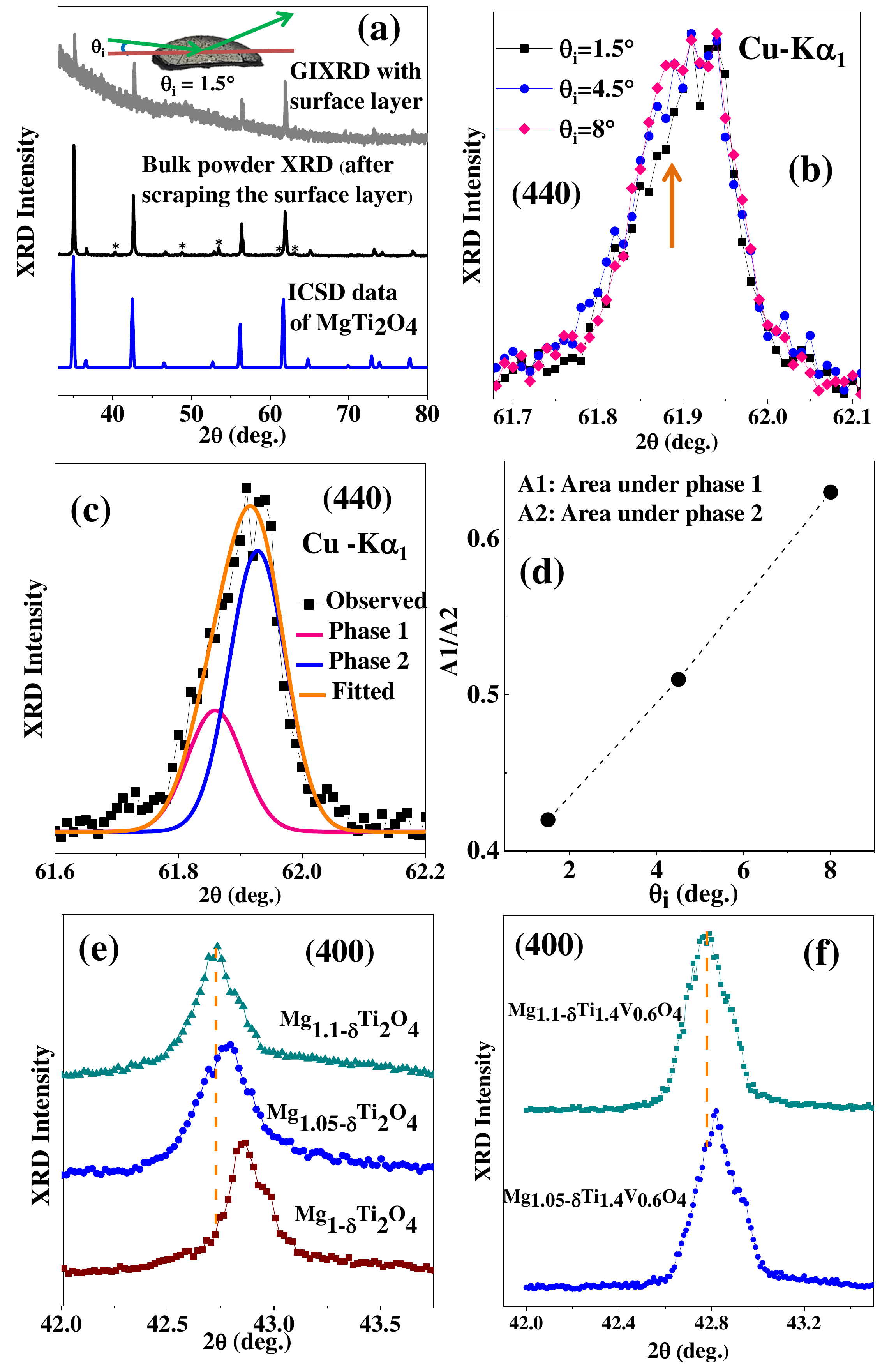}}
		\caption{(a) Comparison of powder XRD of the bulk sample (after scraping of the surface layer), GIXRD for grazing incidence ($\theta_\text{i}$) 1.5$^{\circ}$ with the surface layer  and standard XRD data of bulk MgTi$_2$O$_4$ (obtained from ICSD database). The marked asterisks in the bulk powder XRD arises from a small Ti$_2$O$_3$ secondary phase. (b) Comparison of the (440) GIXRD peak (collected using Ge(220) 2-bounce monochromator) for different grazing incidences. (c) (440) GIXRD peak fitted with two phases. (d) The obtained increase in the ratio of the areas corresponding to phase 1 (A1) and phase 2 (A2) (i.e. $\frac{A1}{A2}$) with increase in the grazing incidence angle, suggesting a relative increase in the phase 1 fraction with increasing sample depth. Shifts of the (400) XRD peaks towards lower 2$\theta$ value with increase in Mg percentage in different (e) MgTi$_2$O$_4$ and (f) V-doped MgTi$_2$O$_4$ samples. 10\% excess Mg was taken to account for Mg volatility following reference \cite{Goodenough2005}.}\label {Structure_sup}
	\end{center}
\end{figure}
\\We first discuss the structural phases for the bulk and the grayish surface layer, as investigated through X-ray diffraction (XRD). As seen through a comparison with XRD diffraction pattern of standard MgTi$_2$O$_4$ in Fig.\ref{Structure_sup}(a), the bulk of the synthesized MgTi$_{1.4}$V$_{0.6}$O$_4$ is found to stabilize into a cubic spinel phase. Along with the main spinel phase, a small fraction of a secondary phase of Ti$_2$O$_3$ (corundum) (the corresponding XRD peaks are indicated by asterisks) can also be detected. Since the Ti$_2$O$_3$ is not superconducting \cite{Morin1959,LHTjeng2018} (also the bulk sample, without the surface layer, is found to be insulating), its presence does not affect the present results. To probe the structural phase of the surface layer, GIXRD with a very low incident grazing angle of 1.5$^{\circ}$ was performed, so that the X-ray beam mostly get diffracted from the surface layer. Clear, though weak (due to low sample volume), characteristic XRD peaks corresponding to two spinel phases, which vary in their lattice parameters (thereby leading to a splitting in the XRD peak positions), can be detected through GIXRD (as seen in Fig.\ref{Structure_sup}(b)). Notably, the GIXRD peaks of the thin grayish surface layer do not match with the XRD pattern corresponding to the Ti$_9$O$_{10}$ orthorhombic structure of the superconducting Mg-Ti-O superconducting films \cite{Jin2022}. The observation of two spinel phases (as seen in Fig.\ref{Structure_sup}(b)) through GIXRD, is in consistence with an inspection of the top grayish surface layer under a microscope (as seen in image of inset of Fig.2), which clearly exhibit two distinct sample regions, i.e. overlapping grayish islands interspersed on relatively blackish sample regions, with the relative content of the later increasing with depth in the sample. To further confirm the two spinel phases, we have collected the GIXRD using Ge (220) 2-bounce monochromator (which suppresses the Cu-K$\alpha_2$ radiation) for different grazing incidences. The intensity of the lower 2$\theta$ peak in GIXRD (as shown in Fig.\ref{Structure_sup}(b) and Fig.\ref{Structure_sup}(d)), gradually increases with increase in incidence angle (which, thereby, probes the structure deeper into the sample) suggests that the higher 2$\theta$ peaks (shown in Fig.\ref{Structure_sup}(c)), associated with a smaller lattice parameter (the lattice parameters obtained from the GIXRD are 8.450 ($\pm$ 0.003)$\AA$ and 8.459 ($\pm$ 0.003)$\AA$ for the surface layer and bulk portion of the sample, respectively), corresponds to the grayish regions of the surface layer. Further, a systematic decrease in the bulk lattice parameter (leading to a tuning of the corresponding XRD peak positions to higher angles, as seen in Figs.\ref{Structure_sup}(e) and (f) on reducing the Mg content in a control Mg$_x$Ti$_2$O$_4$ and V-doped Mg$_x$Ti$_2$O$_4$ series, is clearly observed. The spinel phase corresponding to the higher 2$\theta$ XRD peaks (seen in Fig.\ref{Structure_sup}(c)) is, thus, likely off-stoichiometric (most likely Mg deficient due to its increased volatility at higher sintering temperature), while the spinel phase corresponding to the lower 2$\theta$ XRD peaks is near stoichiometric (comparable to the bulk, as seen in Fig.\ref{super}(a), which is black in colour).  Energy-dispersive x-ray (EDX) analyses of the high-temperature (T$\approx$1080$^{\circ}$C) sintered samples clearly elucidate significant Mg loss from the surface layer and indicate an average composition for the greyish surface regions to be Mg$_{0.22}$Ti$_{0.63}$V$_{1.37}$O$_4$ and the bulk of the sample (obtained after scraping surface layers), corresponds to an average composition of Mg$_{0.91}$Ti$_{1.3}$V$_{0.7}$O$_4$. Stabilization of Mg$_{0.22}$Ti$_{0.63}$V$_{1.37}$O$_4$ as a spinel phase can be understood in light of the multiple investigations that have demonstrated the stability of the AB$_2$O$_4$ spinel structure even in presence of significant A-site vacancies. In Li$_{1-x}$Ti$_2$O$_4$ compounds, Li$_{0.33}$Ti$_2$O$_4$ is known to crystallize in a spinel phase \cite{Attfield2015,Capponi1994}. Interestingly, the superconducting T$_c$ progressively increases with increased Li vacancy ($x$) in Li$_{1-x}$Ti$_2$O$_4$ \cite{Capponi1994}. For Li$_{1-x}$V$_2$O$_4$ compounds, spinel Li$_{0.28}$V$_2$O$_4$ is also found to crystallize in spinel phase \cite{Thackeray1991}. Similarly, the structural stability of spinel Mg$_{1-x}$Al$_2$O$_4$ has also been demonstrated for large $x$ \cite{Giusta1991}. In addition to significant Mg loss, an increase in the relative V content as compared to the Ti ion content in the greyish surface layer spinel phase is likely driven by the additional propensity of V ions than Ti ions to occupy some of the empty 8a A-sites (driven by Mg-ion vacancies), as observed in [(Li$_{0.28}$V$_{0.1}$)$_{8a}$\{V$_{0.4}$\}$_{16c}$[V$_{1.5}$]$_{16d}$O$_4$] \cite{Thackeray1991} and in
[(Li$_{0.33}$)$_{8a}$\{Ti$_{0.56}$\}$_{16c}$[Ti$_{1.44}$]$_{16d}$O$_4$] \cite{Attfield2015}.
\begin{figure}[t!]
	\begin{center}
		\scalebox{1.0}
		{\includegraphics{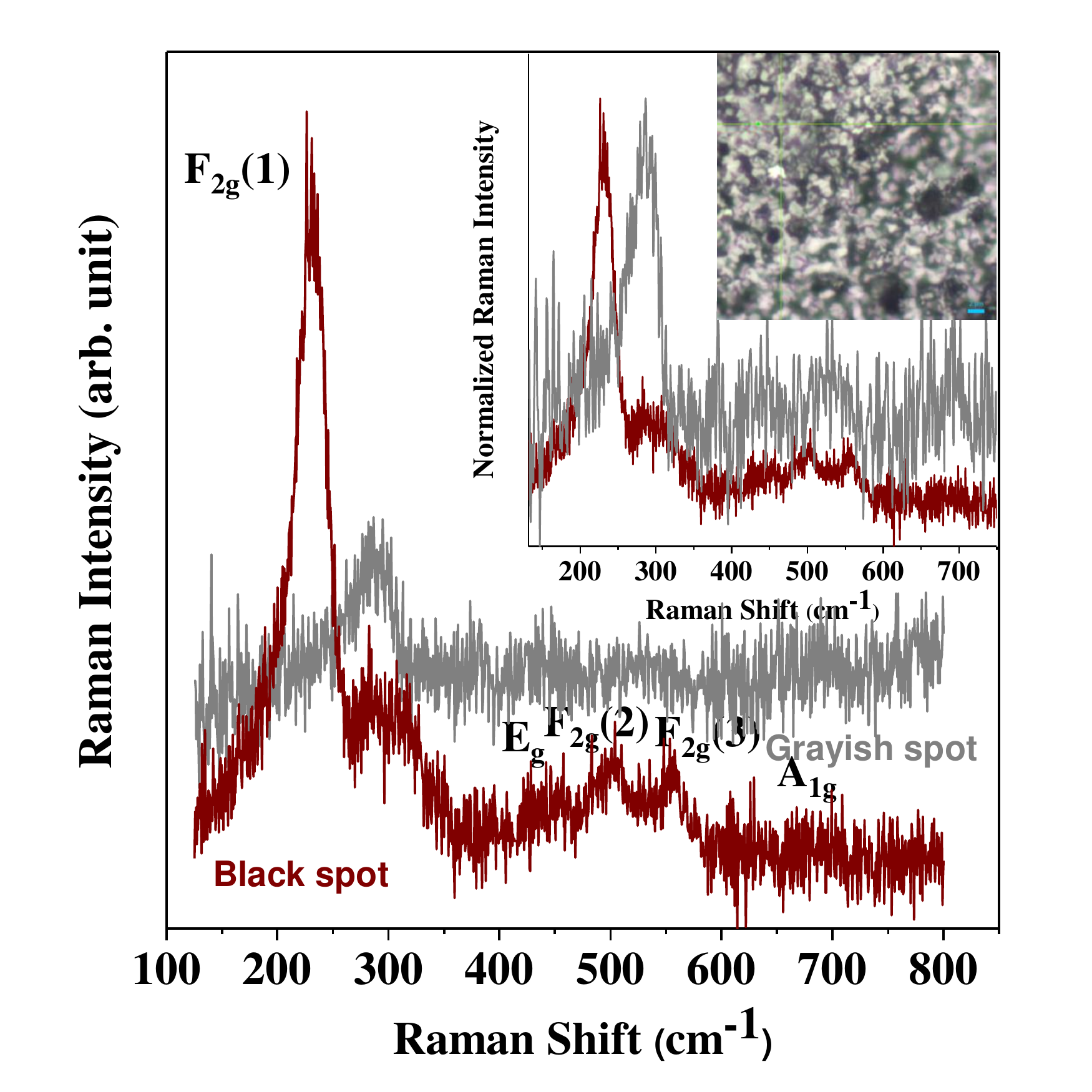}}
		\caption{Comparison of Raman spectra collected on the grayish and black regions of the thin surface layer. Inset shows normalized Raman spectra of both the regions and microscopy image of the grayish surface layer.}\label {Raman}
	\end{center}
\end{figure}
\\To further investigate the structural properties, Micro-Raman measurements were carried out by preliminary focusing the laser beam on the grayish and black regions of the surface layer, as shown in Fig.\ref{Raman}. The corresponding Raman peaks and their positions, the reproducibility of which were checked between different regions, clearly suggest spinel phases for both these regions \cite{Saxena2002,Flor2002,Lottici2015,Gupta2020,Ueda2003}.  The observed main peak around 230 cm$^{-1}$ in the Raman spectra for the black region, is reported to be associated with vibrations involving mainly the $A$O$_4$ (in our case with MgO$_4$) units of the spinel phase \cite{Flor2002,Gupta2020}. The decrease in intensity in the Raman spectra of grayish region suggests the A-site off-stoichiometry and associated disorder for the grayish surface layer. Due to a preponderance of the grayish sample regions over the black sample portions (as seen in inset of Fig.\ref{Raman}) in the top layer (both having typical grain sizes of $\sim$ 1 ${\mu}$m, which is also comparable to the Raman laser-beam spot-size), a small hump, at around 300 cm$^{-1}$, corresponding to the main peak of the gray sample area, becomes discernible in the Raman spectrum collected on the black sample regions, as seen in Fig.\ref{Raman}. However, the shift in the main peak positions (shown in inset of Fig.\ref{Raman}) of the grayish region to higher wavenumber in comparison to the corresponding spectra for the black region indicates an decrease in lattice parameters for the grayish region, in consistence with the GIXRD results.
\begin{figure}[b]
	\begin{center}
		\scalebox{1.2}
		{\includegraphics{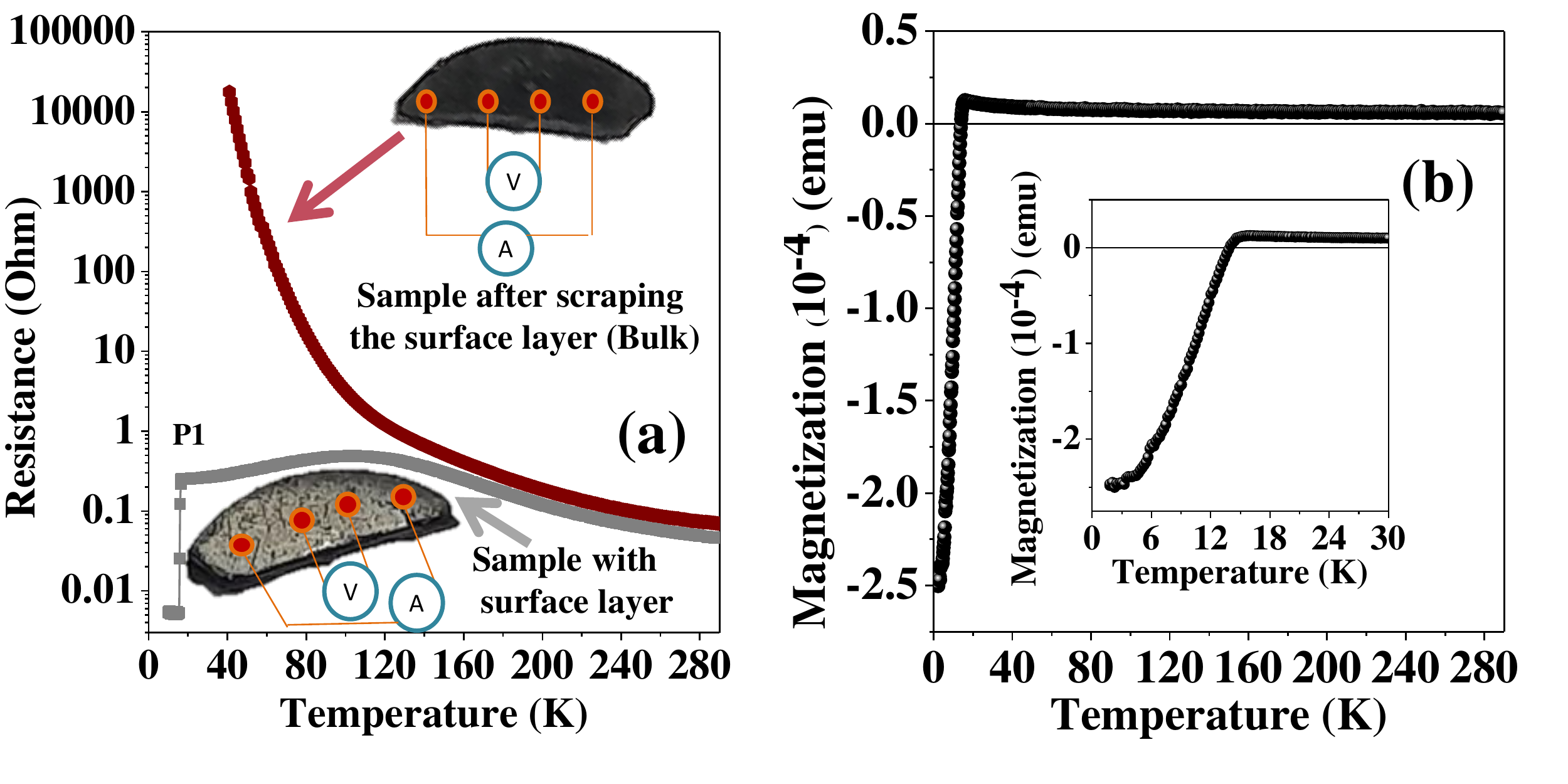}}
		\caption{(a) Comparison of temperature-dependent resistance curves of the sample with and without the grayish surface layer (after scraping the grayish surface layer). The surface of the sample has a different grayish colour (shown in lower inset of the figure) than the blackish bulk (the later picture is taken after scraping the grayish surface layer and is shown in the upper inset of the figure). Temperature dependent resistance curves of the sample with the grayish surface layer exhibits sharp superconducting transitions, while the bulk (after scraping off the grayish surface layer) exhibits semiconducting behaviour.  P1 stands for one piece of sample. (b)Temperature dependence of magnetization, measured with 0.01 Tesla applied magnetic field, collected following zero field cooling protocol. Inset shows zoomed view of the diamagnetic transition below the superconducting transition temperature.}\label {super}
	\end{center}
\end{figure}
\\The temperature-dependent four probe resistivity values of the polycrystalline MgTi$_{1.4}$V$_{0.6}$O$_4$ sample, measured with and without the grayish surface layer, are shown in Fig.\ref{super}(a) (schematic diagram of four probe electrical contacts shown on a real sample image). Surprisingly, while the measurement including the grayish surface layer exhibits a superconducting transition, with a high T$_c$ of around 16 K, the measurement on the sample without the grayish surface layer (i.e. property of the bulk of the sample) leads to an insulating behavior down to the lowest temperature. The high temperature insulating nature which show a similar temperature dependency for both the resistivity curves, appears to be driven by the resistivity of the bulk sample. At temperatures below around 120 K, the transport property of the grayish surface layer seems to dominate over the bulk transport property, suggesting a lower resistance for the surface layer in this temperature range. To further validate the emergence of a superconducting phase within the grayish surface layer, we have measured temperature dependent magnetization measurement on the pellet sample (which included the surface layer). Expectedly, the magnetization curve, as shown in Fig.\ref{super}(b), clearly exhibits a sharp diamagnetic transition below the superconducting transition temperature of around 16 K (seen in the inset of Fig.\ref{super}(b)). Notably, the observed T$_c$ for the superconducting transition is found to be the highest amongst the whole family of superconducting spinel compounds \cite{Nagata1995,Cava2013,Luo2022,Johnston1973,Hitosugi2017}.
\begin{figure}[b]
	\begin{center}
		\scalebox{1.15}
		{\includegraphics{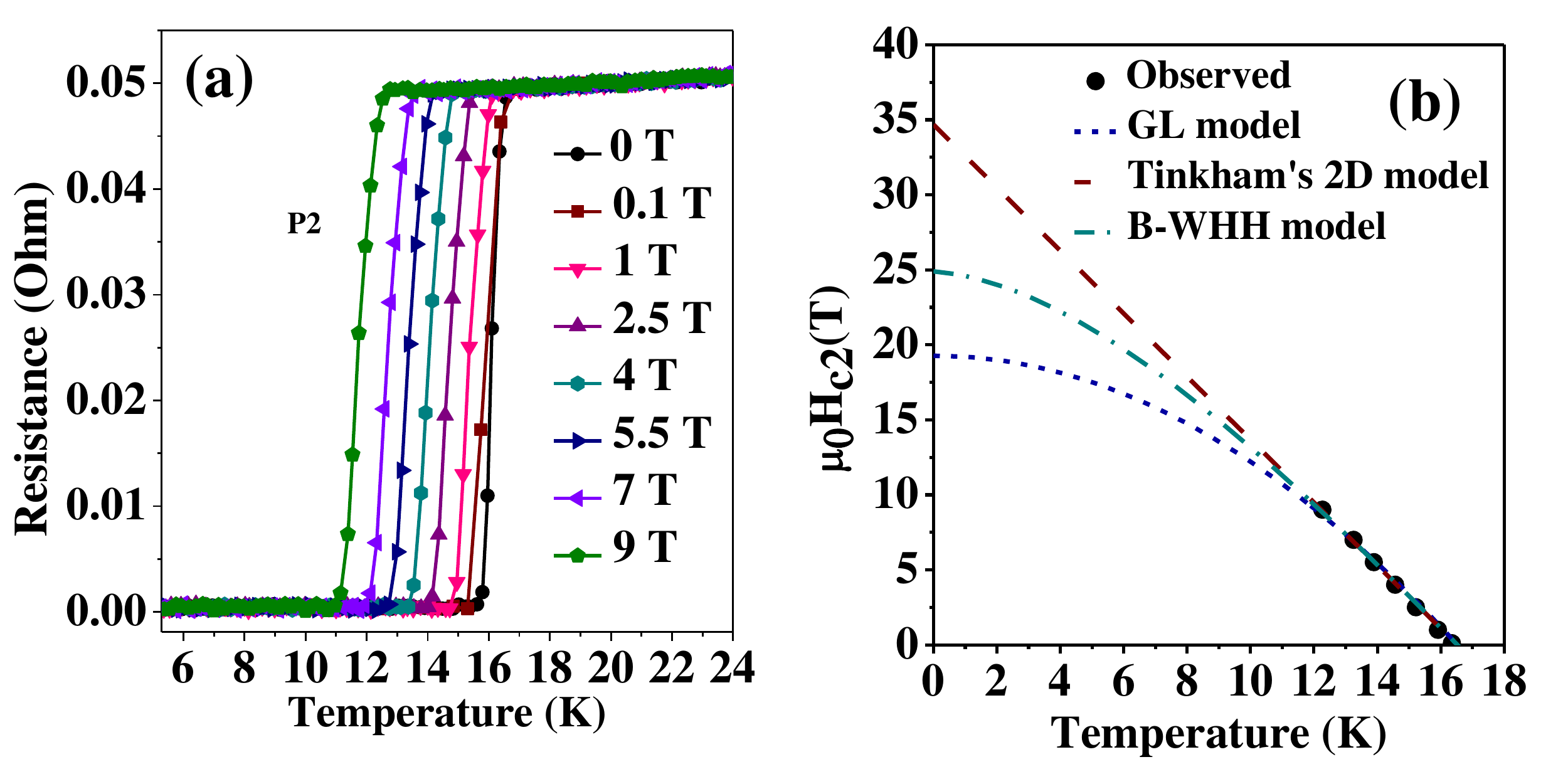}}
		\caption{(a) Temperature dependent resistance curves collected with different applied magnetic fields. P2 stands for another piece of sample. (b) Fitting of the critical magnetic field with superconducting transition temperatures with Ginzburg-Landau (GL), Werthamer-Helfand-Hohenberg (WHH) and Tinkham's 2D model.}\label {super1}
	\end{center}
\end{figure}
\\The temperature dependent resistivity curves, measured with varying applied magnetic fields, illustrates that even a high magnetic field of 9 Tesla remains nearly ineffective in changing the sharpness of the superconducting transition (as seen in Fig.\ref{super1}(a)) or the superconducting T$_c$ substantially (T$_c$ decreases by $\sim$ 4 K for 9 Tesla magnetic field), thereby, suggesting a very high upper critical magnetic field of this system. To estimate the critical magnetic field, the T$_c$ (taken to be the temperature at which the resistance drops to 90\% of the normal state resistance) values corresponding to different magnetic fields have been plotted and fitted with some of the proposed models of superconductivity, such as Ginzburg-Landau \cite{Cyrot1973,Li2010}, Werthamer-Helfand-Hohenberg (WHH) \cite{Hohenberg1966} and Tinkham's 2D model \cite{Tinkham1963,Tinkham1964,Tinkham1968}, as shown in Fig.\ref{super1}(b). The WHH and Tinkham's model, observed to fit the experimental data better in comparison to the Ginzburg-Landau model, suggests a very high upper critical magnetic field, such as $\sim$ 25 Tesla and 35 Tesla, respectively. Further investigations to ascertain the exact critical field value (i.e. to understand whether it is beyond the Pauli paramagnetic limit (B$_{p}$ = 1.84 T$_c$)) will necessiate resistivity measurements with higher magnetic field values. Notably, both the estimated upper critical field values are much higher than those reported for either the sulpho and selenide superconductors (with upper critical magnetic field values less than 5 Tesla \cite{Nagata1995,Cava2013,Luo2022}) or the spinel oxide superconductors (LiTi$_2$O$_4$ and superlattices of MgTi$_2$O$_4$, which have upper critical field values of $\sim$ 12 Tesla \cite{Jin2020,Wu2004,Wen2006,Jin2022}). Thus, the emergence of superconductivity in this system not only leads to the highest T$_c$ among spinel compounds but is also associated with a very high upper critical magnetic field, which is promising. Non-superconducting precipitates (the darker regions of the surface layer), that naturally occur on the surface layer of V- doped Mg$_{1-x}$Ti$_2$O$_4$ along with the superconducting regions, likely acts as very efficient pinning centres for the superconducting vortices. Such pinning centres often are artificially engineered to make the irreversibility magnetic field values (above which the dissipation-less transport or critical-current value vanishes) come close to the critical magnetic field (H$_{c2}$) values, which is essential for applications \cite{Gurevich2011,Lee2009,Wen2008}. The irreversibility magnetic field values of the V-doped Mg$_{1-x}$Ti$_2$O$_4$, as probed using current (I) – voltage (V) characteristics (shown in Fig. \ref{irreversible}(a)) are estimated to be very close to the corresponding H$_{c2}$ values (as shown in Fig. \ref{irreversible}(b) and Fig.\ref{super}(d)), further highlighting the promise of this superconducting system.
\begin{figure}[t]
	\begin{center}
		\scalebox{1.12}
		{\includegraphics{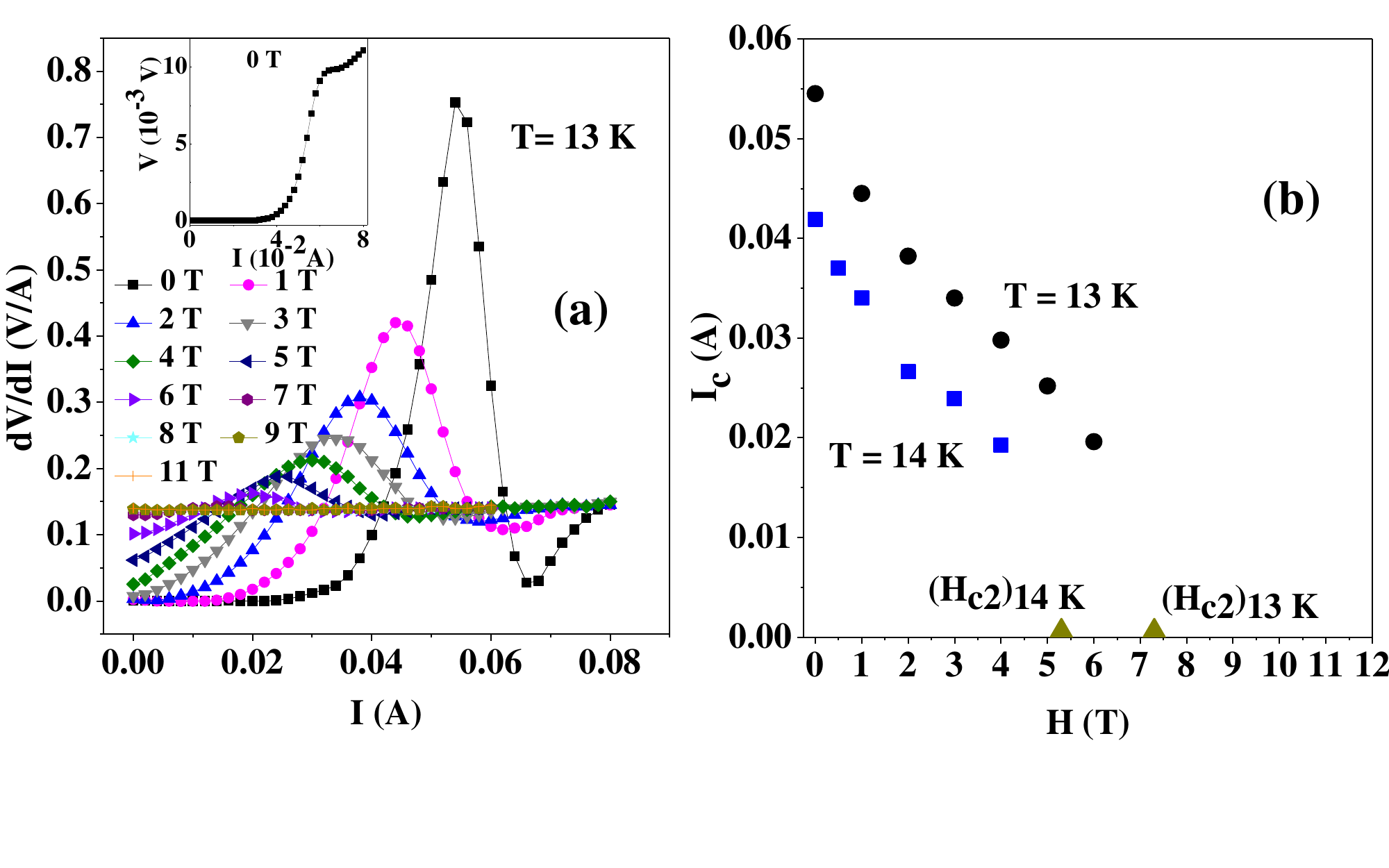}}
		\vspace*{-0.3 in}\caption{(color online) (a) Isothermal $\frac{dV}{dI}$ vs. I curve at 13 K for various applied magnetic field values. Inset shows the corresponding V-I curve in absence of an applied magnetic field. (b) Magnetic field dependence of the critical currents (determined from the current value associated with a peak in the corresponding $\frac{dV}{dI}$ curve, as shown in (a)) at 13 K and 14 K. The upper critical magnetic field (determined from the resistance data of Fig.\ref{super}(d)) corresponding to 13 K and 14 K are also indicated by filled triangles.}\label {irreversible}
	\end{center}
\end{figure}
\\The realization of strained MgTi$_2$O$_4$ in thin-film superlattice with SrTiO$_3$ was reported to be crucial for the onset of superconductivity in MgTi$_2$O$_4$, albeit with a much lower T$_c$ and upper critical magnetic field values of $\sim$ 3 K and 12 Tesla, respectively \cite{Jin2020}.  While a decrease in lattice parameter for the grayish region, aided through Mg-offstoichiometry (which also dopes charge carriers), is naturally realized here (as suggested through GIXRD, Raman and EDAX investigations \cite{Attfield2015,Capponi1994,Thackeray1991,Giusta1991}), the role of V-doping also seems very important. Particularly, as discussed earlier, V-doping into MgTi$_2$O$_4$ does bring in charge and orbital fluctuations, and whether it helps in boosting superconductivity remains to be investigated.
\\In summary, we have reported the emergence of superconductivity on a surface layer of a V-doped MgTi$_2$O$_4$ sample. The superconducting transition temperature and upper critical magnetic field is found to be five times and two times enhanced as compared to MgTi$_2$O$_4$ and SrTiO$_3$ superlattices. The sample off-stoichiometry (Mg deficiency for the spinel phase of the surface layer) along with doping of V ions seem critical for the observed superconductivity.

A.R. would like to acknowledge functional material laboratory members, IIT Kharagpur, for their help and support.
D.C. acknowledges financial support from INSA Young Scientist project Grant No. INSA/SP/YSP/151/2018/233. T.P.
would like to acknowledge SERB for providing support through a fellowship (file No. PDF/2016/002580). A.N.P. acknowledges financial support from DST-Nano Mission Grant No. DST/NM/TUE/QM-10/2019. The authors acknowledge the characterization facilities from the TRC project of SNBNCBS. The authors acknowledge support from Ministry of Electronics and Information Technology (MeitY) No. 5(7)/2017-NANO 5(1)/2021-NANO, and would like to thank
Professor Anushree Roy, Professor Atsushi Fujimori, and Professor Arghya Taraphder for help and fruitful discussions.

\end{document}